\documentclass[preprint2,10pt]{aastex}

\newcommand{\rev}[1]{{#1}}

\begin{document}

\title{
Planetary Core Formation with Collisional Fragmentation and Atmosphere 
to Form Gas Giant Planets
}

\shorttitle{
Formation of Gas Giant Planets 
}
\shortauthors{
Kobayashi, Tanaka, Krivov
}

\author{Hiroshi Kobayashi$^{1}$, Hidekazu Tanaka$^2$, Alexander V.
Krivov$^1$}

\affil{$^{1}$ Astrophysical Institute and University Observatory,
Friedrich Schiller University, Schillergaesschen 2-3, 07745, Jena, GERMANY}

\affil{$^{2}$ Institute of Low Temperature Science, Hokkaido University,
Kita-Ku Kita 19 Nishi 8, Sapporo 060-0819, JAPAN}

\email{hkobayas@astro.uni-jena.de}

\begin{abstract}
 Massive planetary cores ($\sim 10$ Earth masses) trigger rapid gas
 accretion to form gas giant planets \rev{such as} Jupiter and Saturn.  We
 investigate the core growth and the possibilities for cores to reach
 such a critical core mass.  At the late stage, planetary cores grow
 through collisions with small planetesimals.  Collisional fragmentation
 of planetesimals, which is induced by gravitational interaction with
 planetary cores, reduces the amount of planetesimals surrounding them,
 and thus the final core masses.  Starting from small planetesimals that
 the fragmentation rapidly removes, less massive cores are formed.
 However, planetary cores acquire atmospheres that enlarge their
 collisional cross section before rapid gas accretion.  Once planetary
 cores exceed about Mars mass, atmospheres significantly accelerate the
 growth of cores.  We show that, taking into account the effects of
 fragmentation and atmosphere, initially large planetesimals enable
 formation of sufficiently massive cores.  On the other hand, because
 the growth of cores is slow for large planetesimals, a massive disk is
 necessary for cores to grow enough within a disk lifetime.  If the disk
 with 100\,km-sized initial planetesimals is 10 times as massive as the
 minimum mass solar nebula, planetary cores can exceed 10 Earth masses
 in the Jovian planet region ($>5\,$AU).
\end{abstract}

\section{Introduction}

Gas giant planets \rev{such as} Jupiter and Saturn form in gaseous
disks.  In the core-accretion model, the accretion of planetesimals
produces cores of giant planets. Once a core reaches a critical mass
$\sim 10$ Earth masses, it can rapidly accrete gas to form a gas giant
planet \citep{mizuno80,bodenheimer86,ikoma00}. Gas giants must form
within the lifetime of gaseous disks ($\la 10$\,Myr).

For km-sized or larger planetesimals, gravitational focusing enhances
their collisional cross sections, resulting in a high collision 
probability for low relative velocities.  Relative velocities of large
bodies are kept lower than those of small ones due to dynamical 
friction.  A combination of gravitational focusing and dynamical 
friction brings rapid growth of large bodies, which is referred 
to as runaway growth \citep{wetherill89}. Eventually, the 
runaway growth generates a small population of large bodies called 
planetary embryos. Planetary embryos keep their orbital separations and 
hence grow through collisions with surrounding remnant planetesimals 
more slowly than in the runaway mode \citep{kokubo98}. 
This regime is called oligarchic growth.

\citet{kobayashi+10} pointed out that the oligarchic growth halts due to
fragmentation of planetesimals.  In the oligarchic growth, the relative
velocities of planetesimals are controlled by the viscous stirring of
embryos and gas drag. As embryos grow, the velocities of remnant
planetesimals are increased so greatly that collisions between
planetesimals become destructive. Such collisions eject numerous
fragments, which collide with each other to produce further smaller
bodies.  Planetesimals are therefore ground down through such successive
collisions (collision cascade). The random velocities of small bodies
are strongly damped by gas drag and thereby the collisional cascade no
longer occurs for fragments with radii $\la 1$--$10\,{\rm m}$. In the
end such fragments drift inward due to gas drag and are lost around
embryos. The collisional cascade combined with the loss of fragments
reduces the solid surface density and hence final embryo masses.  
\rev{\citet{kobayashi+10} showed that the final embryo masses 
are as small as Mars mass for 1-100\,km-sized planetesimals in
the minimum mass solar nebula \citep[hereafter MMSN;][]{hayashi81}. 
}
Large
planetesimals, which are relatively hard to be broken collisionally,
\rev{and a massive disk} produce massive final embryos.  \rev{However, }
collisional fragmentation makes it
difficult to form giant planets along the lines of the core-accretion
model; starting from 100\,km-sized planetesimals, planetary embryos can
reach the critical core mass for gas accretion only inside 3--4\,AU in a
disk that is 10 times more massive than the MMSN model. 

The motion of fragments $\la 1$\,m is coupled with gas.  The drift
timescale of such fragments are relatively long. 
\citet{kenyon09} proposed that embryos may accrete a large amount of such
fragments. However, the strong gas drag in the Stokes regime is dominant
for fragments $\la 100$\,m and damps the relative velocities to halt
collision cascade at $1$--$10\,$m as mentioned above. Therefore, only a
small amount of coupled bodies are produced and hence they hardly
contribute to embryo growth \citep{kobayashi+10}.

Many authors have investigated embryo growth with $N$-body, statistical,
and hybrid simulations
\citep{kokubo96,kokubo98,kokubo00,kokubo02,inaba99,inaba01,inaba03,weidenschilling97,weidenschilling05,weidenschilling08,kenyon04,kenyon08,chambers06,chambers08,kobayashi+10}.
Although providing most accurate dynamical results, $N$-body simulations
have difficulty in producing numerous fragments and following their
fate.  The fragmentation effect on embryo growth has thus not been
treated in detail in spite of its importance.  Recently, Levison et
al. (2010) included fragment production in their $N$-body
simulation. However, it is still difficult to treat fragment--fragment
collisions.  Such successive collisions are essential in the collision
cascade \citep[e.g.,][]{kobayashi10}.  Therefore, statistical
simulations are a better method to accurately investigate planet
formation with fragmentation.

In the statistical simulation, the collisional mass evolution of bodies
is calculated within a ``particle-in-a-box'' approximation.  Bodies have
horizontal and vertical components of random velocity relative to a
circular orbit that are determined by their eccentricities and
inclinations, respectively.  These velocities are changed by
gravitational interactions between the bodies and hence affected by
their mass spectrum, while the collision rates between the bodies depend
on the velocities. Therefore, the coupled mass and velocity evolution
needs to be solved \citep{wetherill93}.  While the statistical method
has advantages, its weak point is the inability to track the individual
positions of planetesimals.  However, progress in planetary dynamic
theory \citep{greenzweig92,ida89,Ohtsuki99,stewart00,Ohtsuki02}
has helped to overcome
this problem. For example, \citet{greenzweig92} and \citet{ida89} provided detailed expressions for the probability of
collisions between planetesimals orbiting a central star, while \cite{stewart00} and \citet{Ohtsuki02} derived improved equations for
calculating the evolution of random planetesimal velocities caused by
gravitational interactions.  Finally, it has been shown that the
recently developed statistical codes can describe some aspects of the
planetary accumulation processes with the same accuracy as $N$-body
simulations \citep{inaba01,kobayashi+10}. 


Since the timescale of collision cascade strongly affects the final mass
of planetary embryos \citep{kobayashi+10}, fragmentation outcome
models are essential for embryo growth.  Collisional fragmentation
includes several uncertain parameters.  \citet{kobayashi10}
constructed a simple fragmentation model which is consistent with
laboratory experiments \citep{fujiwara,takagi,holsapple} and
hydrodynamical simulations \citep{benz99} and analytically
clarified which parameters are essential.  They found that the mass
depletion due to collision cascades is sensitive to the total ejecta
mass yielded by a single collision, while it is almost independent of
the mass of the largest ejecta fragment and the size distribution of
ejecta over a realistic parameter region.  Furthermore, fragmenting
collisions are subdivided into two types, catastrophic disruption and
cratering (erosive collision).  Although some studies neglected or
underestimated the effect of cratering
\citep{dohnanyi,williams,wetherill93,inaba03,bottke}, Kobayashi \&
Tanaka showed that cratering collisions make a dominant contribution to
the collision cascade.

A planetary embryo larger than $\sim 10^{-2}$ Earth masses acquires a
tenuous atmosphere of gas from the disk.  Fragments are captured by the
atmosphere even if they do not collide directly with the embryo,
implying that the collisional cross section of the embryo is enhanced.
This effect advances the growth of Mars-mass or larger embryos
\citep{inaba_ikoma03}.  Embryos with the atmospheres accrete
fragments prior to their drift inward and can acquire more than $10$
Earth masses starting from 10km-sized planetesimals \citep{inaba03}.  However, erosive collisions and initial planetesimal sizes
strongly affect final embryo masses \citep{kobayashi+10}.

This paper investigates the embryo growth taking into account erosive
collisions and embryo's atmosphere.  Although growing embryos may fall
into a central star due to the type I migration
\citep[e.g.,][]{tanaka02}, we neglect the migration here.  We perform
both analytical studies and statistic simulations, which extend those of
Kobayashi et al. (2010) by including atmospheric enhancement of embryo
growth. The goal is to find out what determines embryo growth and
whether an embryo can reach the critical core mass.  We introduce the
theoretical model in Section \ref{sec:theo} and derive final embryo
masses taking into account atmosphere in Section \ref{sc:final_mass}. In
Section \ref{sc:simulation}, we check solutions for final masses against
the statistical simulations. Sections \ref{sc:discussion} and
\ref{sc:summary} contain a discussion and a summary of our findings.

\section{THEORETICAL MODEL}
\label{sec:theo}

\subsection{Disk Model}

We introduce a power-law disk model for the initial surface mass density
of solids $\Sigma_{\rm s,0}$ and gas $\Sigma_{\rm g,0}$ such that
\begin{eqnarray}
 \Sigma_{\rm s,0} &=& f_{\rm ice} \Sigma_1 \left(\frac{a}{1{\rm AU}}\right)^{-q}
  \, {\rm g\, cm}^{-2}, \\
\Sigma_{\rm gas,0} &=& f_{\rm gas}  \Sigma_1 \left(\frac{a}{1{\rm
					  AU}}\right)^{-q} 
  \, {\rm g\, cm}^{-2}, 
\end{eqnarray}
where $a$ is a distance from a central star, $\Sigma_1$ is the reference
surface density at 1\,AU, and $q$ is the power-law index of the radial
distribution.  The gas-dust ratio $f_{\rm gas} = 240$ (Hayashi 1981).
The factor $f_{\rm ice}$ that represents the increase of solid density
by ice condensation beyond the snow line $a_{\rm ice}$ is given by
$f_{\rm ice} = 1$ ($a<a_{\rm ice}$) and 4.2 ($a \geq a_{\rm ice}$).  In
the MMSN model, $\Sigma_1 = 7.1 \,{\rm g\,cm}^{-2}$ and $q = 3/2$.  If
the disk is optically thin,
\begin{equation}
a_{\rm ice} = 2.7 \left(\frac{L_*}{L_\sun}\right)^{1/2} {\rm AU},\label{eq:ice_line} 
\end{equation}
where $L_*$ and $L_\sun$ are the luminosities of the central star and
the sun, respectively.  In reality, disks may be optically thick even
after planetesimal formation. However, we assume
Equation~(\ref{eq:ice_line}) for simplicity.

\subsection{Fragmentation Outcome Model
\label{sc:outcome_model}}

\citet{kobayashi10} showed erosive collisions to dominate the collision
cascade. We should take into account such collisions properly.  We
assume that fragmentation outcomes are scaled by the impact energy and
hence the total ejecta mass $m_{\rm e}$ produced by a single collision
between $m_1$ and $m_2$ is given by a function of the dimensionless
impact energy $\phi = m_1 m_2 v^2/ 2 (m_1+m_2)^2 Q_{\rm D}^*$, where $v$
is the collisional velocity between $m_1$ and $m_2$ and $Q_{\rm D}^*$ is
the specific energy needed for $m_{\rm e} = (m_1 + m_2)/2$. Following
\citet{kobayashi10} and \citet{kobayashi+10}, we model
\begin{equation}
 \frac{m_{\rm e}}{m_1+m_2} = \frac{\phi}{1+\phi}. \label{eq:fragmass}  
\end{equation} 

\begin{figure}[htb]
\epsscale{0.9} \plotone{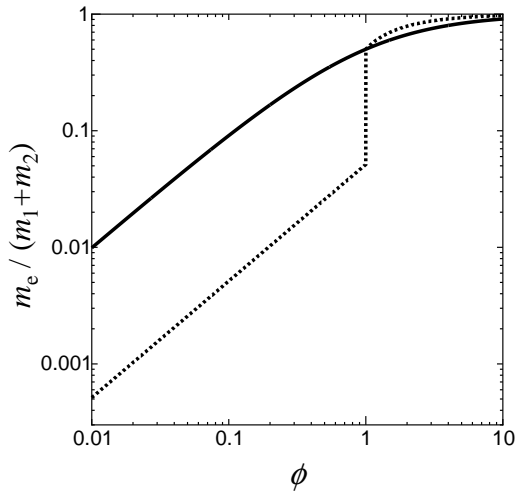} \figcaption{ The total ejecta
mass $m_{\rm e}$ produced by a single collision with $m_1$ and $m_2$ ,
as a function of the dimensionless energy $\phi = m_1 m_2 v^2/ 2
(m_1+m_2)^2 Q_{\rm D}^*$.  The solid line indicates
Equation~(\ref{eq:fragmass}).  For reference, the dotted lines are shown
for the fragment model of \citet{inaba03} with $m_1=10^3\, m_2 =
4.2\times 10^{20}$\,g.  \label{fig:me}}
\end{figure}

\citet{inaba03} derived $m_{\rm e}$ from the fragment model developed by
\citet{wetherill93} with a value of $Q_{\rm D}^*$ found by
\citet{benz99} for ice.  Fig.~\ref{fig:me} shows their model and
Equation~(\ref{eq:fragmass}).  {As discussed in \citet{kobayashi10}},
most of the laboratory experiments and the hydrodynamic numerical
simulations of collisional disruption showed $m_{\rm e}$ not to have a
discontinuity at $\phi = 1$ \citep{housen,takagi,benz99}.  Therefore,
Equation~(\ref{eq:fragmass}) includes erosive collisions ($\phi < 1$)
more accurately.

The critical energy $Q_{\rm D}^*$ is given by 
\begin{eqnarray}
\displaystyle
 Q_{\rm D}^* = Q_{\rm 0s} \Biggl(\frac{r}{1\,{\rm cm}}\Biggr)^{\beta_{\rm s}} +
Q_{\rm 0g} \rho_{\rm p} \Biggl(\frac{r}{1\,{\rm cm}}\Biggr)^{\beta_{\rm g}} +
C_{\rm gg}
\frac{2Gm}{r}
,\label{eq:qd} 
\end{eqnarray}
where $r$ and $m$ are the radius and mass of a body, $\rho_{\rm p}$ is
its density, and $G$ is the gravitational constant.  The first term on
the right-hand side of Equation~(\ref{eq:qd}) is dominant for $r \la
10^4$--$10^5$\,cm, the second term describes $Q_{\rm D}^*$ of $r \la
10^7$\,cm, and the third term controls $Q_{\rm D}^*$ for the larger
bodies.  \citet{benz99} performed the hydrodynamical simulations of
collisional dispersion for $r = 1$--$10^7\,$cm and provided the values
of $Q_{\rm 0s}, {\beta_{\rm s}}, Q_{\rm 0g}$, and ${\beta_{\rm g}}$.
For $r \ga 10^7$\,cm, $Q_{\rm D}^*$ is purely determined by the
gravitational binding energy, being independent of material properties.
The collisional simulation for gravitational aggregates yields $C_{\rm
gg} \sim 10$ \citep{stewart09}.


\subsection{Enhancement Radius by Atmosphere
\label{sc:enhance_rad}
}
Once a planetary embryo has grown larger than the Moon, it acquires
an atmosphere.  It helps the accretion of planetesimals or fragments onto
an embryo; small bodies are captured by the atmosphere of the embryo. 

\citet{inaba_ikoma03} provided an analytical model for a density profile
of the atmosphere.  We consider the atmosphere at a distance $R_{\rm e}$
from an embryo center.  We assume that $R_{\rm e}$ is much smaller than
that at the outer boundary of the atmosphere and that its temperature is
much higher than that at the boundary.  The atmospheric density
$\rho_{\rm a}$ is then proportional to $R_{\rm e}^{-3}$ 
\citep{mizuno80,stevenson82}. Applying the temperature $T_{\rm neb}$, pressure
$P_{\rm neb}$, and density $\rho_{\rm neb}$ of the nebula in the disk
midplane as those at the outer boundary of atmosphere, the density
profile of the atmosphere around an embryo with mass $M$ is given by
\begin{equation}
 \frac{\rho_{\rm a}(R_{\rm e})}{\rho_{\rm neb}} = \frac{16 \pi \sigma_{\rm SB} G M
  T_{\rm neb}^4 }{3 \kappa L_{\rm e} P_{\rm neb}}
  \left(
   \frac{GM\rho_{\rm neb}}{4 P_{\rm neb} R_{\rm e}}
\right)^3,\label{eq:atm_dens} 
\end{equation}
where $\kappa$ is the opacity of the atmosphere and $\sigma_{\rm SB}$ is
the Stephan-Boltzmann constant.  The planetary luminosity $L_{\rm e}$
mainly comes from the accretion of bodies. We approximate
\begin{equation}
 L_{\rm e} = \frac{GM}{R}\frac{dM}{dt},
\end{equation}
where $R$ is the embryo radius.  To validate the assumption of
$\rho_{\rm a} \propto R_{\rm e}^{-3}$, we will apply the complete model
by \citet{inaba_ikoma03} to our statistic simulation and compare our
analytical solutions with the statistical simulations in Section
\ref{sc:simulation}.

When a body passes by a planetary embryo with an atmosphere, the
embryo can accrete the body without direct collision due to the
atmosphere.  The relative velocity between the body and the embryo at
infinity is determined by the eccentricity $e$ of the small body; it is
given by $e v_{\rm k}$ with the Keplerian velocity $v_{\rm k} =
\sqrt{GM_*/a}$ \rev{and} $M_*$ being the mass of a central star.  The relative
velocity is typically smaller than the surface escape velocity of the
embryo during the embryo growth.  If the orbital energy of a body is
sufficiently reduced by the atmospheric gas drag, the body is captured
by the embryo.  The maximum radius $r$ of bodies captured at distance
$R_{\rm e}$ is given by \citep{inaba_ikoma03}
\begin{equation}
 r = \frac{9 a h_{M}}{6+ \tilde e^2} \frac{\rho_{\rm
  a}(R_{\rm e})}{\rho_{\rm p}},\label{eq:atm_cap} 
\end{equation}
where $h_M = (M/3M_*)^{1/3}$ is the reduced Hill radius of the embryo
and $\tilde e = e/h_M$.  Equation~(\ref{eq:atm_cap}) is derived under
the two-body approximation.  \citet{tanigawa10} confirmed that
Equation~(\ref{eq:atm_cap}) is valid in the case where the three-body
effects are included.

Equation~(\ref{eq:atm_cap}) means that $R_{\rm e}$ is the effective
collisional radius of an embryo for bodies with radius $r$.  The
enhanced radius of the embryo with atmosphere is thus derived from
Eqs.~(\ref{eq:atm_dens})--(\ref{eq:atm_cap}) as
\begin{equation}
 \frac{R_{\rm e}}{R} = \frac{F M^{8/9}}{m^{1/9}\dot M^{1/3}}, \label{eq:Re} 
\end{equation}
where 
\begin{equation}
F = \left[
\frac{\pi^2 a \sigma_{\rm SB} T_{\rm neb}^4\rho_{\rm neb}^4 G^3}{(\tilde e^2+6) (3 M_*)^{1/3} \kappa P_{\rm neb}^4}
\right]^{1/3}. 
\end{equation}

\begin{figure}[htb]
\epsscale{0.9} \plotone{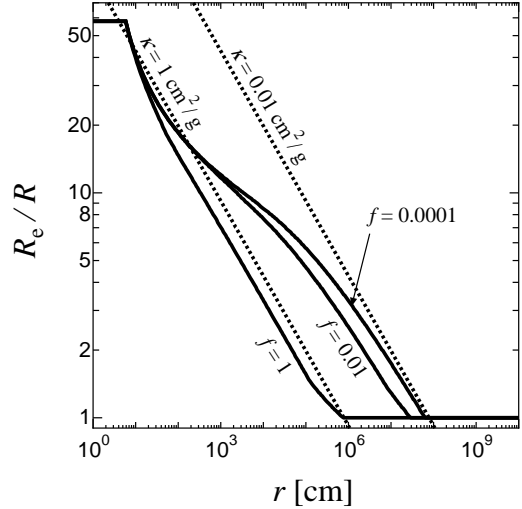} \figcaption{ 
The ratio of the enhanced radius of planetary embryo to its physical radius 
with $M= M_\oplus$, $\rho_{\rm p} = 1 {\rm g \, cm}^{-3}$, and $\dot M = 1\times 10^{-6} M_\oplus/$yr for $\tilde e =
 4$ \rev{in the MMSN disk around the star with mass $M_\sun$}. The ratios are calculated by
 the formulae of \citet{inaba_ikoma03} for \rev{the opacity obtained from
 Equation (\ref{eq:kappa}) with the grain depletion factor} $f=10^{-4}$--1 (solid
 lines) and by Equation~(\ref{eq:Re}) for \rev{the constant opacity} $\kappa = 0.01\,{\rm
 cm}^2\,{\rm g}^{-1}$ and 
 $1\,{\rm cm}^2\,{\rm g}^{-1}$ (dotted lines). 
\label{fig:enhanced_radius}}
\end{figure}

The enhancement factor $R_{\rm e}/R$ given by Equation~(\ref{eq:Re}) is
shown in Fig.~\ref{fig:enhanced_radius}, where the power-law density
profile given by Equation~(\ref{eq:atm_dens}) is compared with a more
realistic profile given by \citet{inaba_ikoma03}.  As we discuss later,
planetary embryos mainly grow through collisions with planetesimals of
the initial size or with fragments of radius $r \sim 10$\,m.  The
enhancement factor calculated with Equation~(\ref{eq:Re}) reproduces
well the more realistic one for km-sized or larger planetesimals, but
Equation~(\ref{eq:Re}) significantly overestimates $R_{\rm e}/R$ for
fragments.  However, since the accretion rate due to collision with such
fragments has a weak dependence on the enhancement factor ($\propto
(R_{\rm e}/R)^{1/2}$; see Equation~(\ref{eq:pcol_low})), this
discrepancy produces insignificant errors.

\section{FINAL EMBRYO MASS  
}
\label{sc:final_mass}

\subsection{Isolation Mass}

Planetary embryos can grow until they have accreted all planetesimals
within their feeding zones. The width of a feeding zone is given by the
orbital separation of \rev{neighboring} embryos, $\tilde b
(2M/3M_*)^{1/3} a$, where $\tilde b \simeq 10$ is \rev{the separation measured in their mutual
Hill radii} \citep{kokubo00,kokubo02}.  The maximum
mass or ``isolation mass'' is $M_{\rm iso} = 2 \pi a^2 (2M_{\rm
iso}/3M_*)^{1/3}\tilde b \Sigma_{\rm s,0}$.  It can be expressed as
\begin{eqnarray}
 M_{\rm iso} &=& 2.8 \Biggl(\frac{\tilde b}{10}\Biggr)^{3/2}
  \Biggl(\frac{\Sigma_{\rm s,0}}{2.7\,{\rm g\,cm}^{-2}}\Biggr)^{3/2}
  \nonumber 
  \\&& \times 
  \Biggl(\frac{a}{5\,{\rm AU}}\Biggr)^3
  \Biggl(\frac{M_*}{M_\sun}\Biggr)^{-1/2} M_{\oplus},\label{eq:M_iso} 
\end{eqnarray}
where $M_{\oplus}$ is the Earth mass and $M_\sun$ is the solar mass.
The planetary embryo mass approaches the isolation mass if fragmentation
is ignored \citep{kokubo00,kokubo02}.  However, if fragmentation is
included, the embryo mass can reach only about Mars mass for a MMSN disk
\citep{kobayashi+10}.

\subsection{Planetesimal Accretion}

As shown by \citet{kobayashi+10}, a planetary embryo accretes
planetesimals with masses comparable to original ones or fragments
resulting from collisional grinding of planetesimals.  In the former
case, a final embryo mass is determined by the equilibrium between the
accretion of planetesimals and their removal due to collisional
grinding. In the latter case, an embryo can grow until fragments are
depleted by the gas drag. Following \citet{kobayashi+10}, we want
here to derive final masses determined by the accretion of planetesimals 
in the case with atmospheric enhancement, while we treat the fragment
accretion in Section \ref{sc:fragment}. 

At the oligarchic stage, embryos mainly grow through collisions with
planetesimals that dominate the surface density. 
The growth rate of an embryo with mass $M$ is given by 
\begin{equation}
 \frac{dM}{dt} = C_{\rm acc} \Sigma_{\rm s} a^2 h^2_{M} \langle P_{\rm
  col} \rangle \Omega_{\rm k},\label{eq:dM_ori} 
\end{equation}
where $\Omega_{\rm K}$ is the Keplerian frequency and $C_{\rm acc}$ is
the correction factor on the order of unity.  The dimensionless
collision rate $\langle P_{\rm col} \rangle$ is formulated as a function
of the eccentricities $e$ and inclinations $i$ of bodies accreted onto the
embryo. We assume $e=2i$ in this analysis. 

Embryos have a constant ratio of their separations to their Hill
radii \citep{kokubo98}.  When the ratio decreases as embryos grow,
relatively smaller embryos are culled and thereby remaining embryos 
keep the ratio constant.  Supposing the cull occurs instantaneously, the 
growth rate of embryos due to the cull is estimated to be a half of that from
planetesimals. Therefore, we set $C_{\rm acc} = 1.5$ 
\rev{\citep[e.g.,][]{chambers06,kobayashi+10}.}

For kilometer-sized or larger planetesimals,
their eccentricities $e$ are controlled by the
embryo stirring and gas drag. 
\rev{
The stirring rate is written as $de^2/dt = n_M a^2 h_M^4 \langle P_{\rm
VS} \rangle \Omega_{\rm k}$, where $n_M$ is the surface number density of
embryos and the dimensionless stirring rate $\langle P_{\rm VS} \rangle$
is given by $\langle P_{\rm VS} \rangle = C_{\rm VS} h_M^2 \ln
(\Lambda^2+1)/e^2$ with $C_{\rm VS} = 40$ and $\Lambda = 5 \tilde
e^3/96$ for $e \gg h_M$ \citep{Ohtsuki02}. 
Although $\ln (\Lambda^2+1)$ in $\langle P_{\rm VS} \rangle$ is weakly dependent on $e$, we adopt, in
this analysis, $\ln (\Lambda^2+1) \simeq 8$, with which value we can 
reproduce the formula of \citet{Ohtsuki02} for $\tilde e =
3$--10. 
The gas-drag time $\tau$ is characterized as \citep{adachi76}
\begin{equation}
 \tau = \frac{2 m}{\pi r^2 C_{\rm D} \rho_{\rm neb} v_{\rm k}},\label{eq:tau}  
\end{equation}
where the dimensionless gas drag coefficient $C_{\rm D}= 0.5$ for
km-sized or larger planetesimals. It should be noted that $\tau$ is the
stopping time due to gas drag only when the relative velocity $u$
between gas and a body is equal to the Keplerian velocity; hence 
$\tau$ is almost always much shorter than the stopping time for
realistic relative velocities.
The $e$-damping rate due to gas drag is given by $de^2/dt = - C_{\rm
gas} e^3/\tau$ with $C_{\rm gas} = 2.1$ \citep{inaba01}. 
Using $n_{M} = ( 2 \pi a \delta a)^{-1}$ with the orbital separation of
neighboring embryos of $\delta a = 2^{1/3} h_M a \tilde b$
\citep{kokubo00} and equating the stirring and damping
rates result in the equilibrium
eccentricity:} \footnote{\rev{\citet{ida93} and \citet{thommes03} presented
a similar equation from the stirring timescale derived by
\citet{ida93}. We apply the formula of \citet{Ohtsuki02}, which weakly
depends on $e$ through $\ln (\Lambda^2+1)$. 
However, since we adopt a constant value for $\ln (\Lambda^2+1)$ in this
analysis, there is no substantial difference between their and our treatment, except for the
definition of the coefficient for the viscous stirring. }
}
\begin{equation}
 \tilde e = \left[
\frac{C_{\rm VS} \ln (\Lambda^2+1) \Omega_{\rm k} \tau}{2^{4/3} \pi
\tilde b C_{\rm gas}}
\right]^{1/5}. \label{eq:e_pm} 
\end{equation}
Since we roughly estimate $\tilde e \sim (\tau \Omega_{\rm k})^{1/5}$
from Equation~(\ref{eq:e_pm}), the eccentricities of the kilometer-sized
and larger bodies are larger than $h_{M}$\rev{, according to the
assumption $e \gg h_M$. }

Taking into account the
enhancement due to the atmosphere, \rev{
the dimensionless collisional probability for $e \gg h_M$ is given by }
\citep{greenzweig92,inaba01,inaba_ikoma03}
\begin{equation}
 \langle P_{\rm col} \rangle = \frac{C_{\rm col} \tilde R}{\tilde e^2} 
  \frac{R_{\rm e}}{R},\label{eq:pcol_p} 
\end{equation}
where $C_{\rm col} = 36$ and $\tilde R = R/a h_M = (9M_*/4\pi\rho_{\rm
p})^{1/3}/a$.  Inserting Eqs.~(\ref{eq:Re}) and (\ref{eq:pcol_p}) to
Equation~(\ref{eq:dM_ori}), we obtain $\dot M$ as
\begin{equation}
 \frac{dM}{dt} = A_{\rm ca} M^{7/6} \Sigma_{\rm s}^{3/4},\label{eq:dM_pla} 
\end{equation} 
where 
\begin{equation}
 A_{\rm ca} = \left[
\frac{C_{\rm acc} a^2 C_{\rm col} \tilde R F \Omega_{\rm
k}}{(3M_*)^{2/3} \tilde e^2 m^{1/9}}
\right]^{3/4}. 
\end{equation}

\rev{As embryos grow, destructive collisions between planetesimals are
induced by the stirring of embryos 
and generate a lot of small fragments, which produce further small
bodies through mutual collisions. }
Since \rev{very} small bodies resulting from successive collisions are rapidly
removed by the gas drag, {\rm the} collision cascade reduces the surface density
of solids.  
\rev{In the collision cascade, 
collisional fragmentation dominates 
the mass flux along the mass coordinate. 
Since the mass flux is independent of mass in a steady state, 
the mass distribution of fragments 
follows a power law and the power-law exponent
$\alpha$ is given by $\alpha = (11+3p)/(6+3p)$ for $e^2/Q_{\rm D}^*
\propto m^{-p}$ \citep{kobayashi10}. The steady-state mass flux
determines the surface density reduction as}
\citep{kobayashi10,kobayashi+10}
\begin{eqnarray}
 \frac{d \Sigma_{\rm s}}{dt} &=&
  - B_{\rm ca} \Sigma_{\rm s}^2 M^{2(\alpha-1)/3}. 
\label{eq:dsigma_dt} 
\\
B_{\rm ca} &=& \frac{(2-\alpha)^2
  \Omega_{\rm k} s_{123}(\alpha)}{m^{1/3}} 
\nonumber\\ && \times 
\left(\frac{\tilde e^2 v_{\rm k}^2}{2
			    (3M_*)^{2/3} Q_{\rm
			    D}^*}\right)^{\alpha-1},
\end{eqnarray}
where  
\begin{eqnarray}
 s_{123}(\alpha) &=& \int_0^\infty 
\left[\frac{\phi}{2-b} -\phi \ln \frac{\epsilon \phi}{(1+\phi)^2} 
 +\ln(1+\phi)\right]
\nonumber
\\&& \quad \quad 
\times \frac{\phi^{-\alpha}}{1+\phi} d \phi, 
\end{eqnarray}
and $h_0=1.1\rho_{\rm p}^{-2/3}$. For the derivation of Equation~(\ref{eq:dsigma_dt}),
we apply the fragmentation outcome model of \citet{kobayashi10}; ejecta
yielded by a single collision between $m_1$ and $m_2$ are characterised
by their total mass $m_{\rm e}$ and their power-law mass spectrum with
an exponent $b$ below the mass $m_{\rm L} = \epsilon
(m_1+m_2)\phi/(1+\phi)^2$, where $\epsilon < 1$ is a constant.  The
$\Sigma_{\rm s}$ reduction rate is insensitive to $\epsilon$ and $b$
\citep{kobayashi10}. We set $b = 5/3$ and $\epsilon=0.2$ in this paper.

Dividing Equation~(\ref{eq:dM_pla}) by Equation~(\ref{eq:dsigma_dt}) and
integrating, we obtain the relation between the embryo mass $M$ and the
surface density $\Sigma_{\rm s}$:
\begin{eqnarray}
 \frac{6}{4\alpha-5} \left[M^{(4\alpha-5)/6}- M_0^{(4\alpha-5)/6}\right]
\nonumber
\\
  = 4 (\Sigma_{\rm s}^{-1/4} -\Sigma_{\rm s,0}^{-1/4}) \frac{A_{\rm ca}}{B_{\rm ca}} ,\label{eq:mass_sigma} 
\end{eqnarray}
where $M_0$ is the initial embryo mass.  
\rev{
Note that the derivation of Equation (\ref{eq:mass_sigma}) assumed that
the planetesimal density reduction is caused by collisional grinding, 
but 
the planetesimal accretion onto embryos significantly contributes to the
$\Sigma_{\rm s}$-reduction when 
the surface density of
planetesimals, $\Sigma_{\rm s}$, is much smaller
than that of embryos, $M n_M$. 
} When
an embryo reaches a final mass $M_ {\rm ca}$, \rev{$\Sigma_{\rm s}$ may be
described as $C_{\rm \Sigma_{\rm s}} M_{\rm ca} n_M$ with a constant
$C_{\Sigma_{\rm s}} \ll 1$; hence }
\begin{equation}
 \frac{\Sigma_{\rm s}}{\Sigma_{\rm s,0}} = C_{\Sigma_{\rm s}} 
\left(
\frac{M_{\rm ca}}{M_{\rm iso}} 
\right)^{2/3}. \label{eq:sigma_limit} 
\end{equation}
\rev{For 
$C_{\Sigma_{\rm s}} \sim 0.1$, a final mass is consistent with 
simulations \citep{kobayashi+10}. }
We \rev{thus} set $C_{\Sigma_{\rm s}} = 0.1$ to derive a final mass.
From Eqs. (\ref{eq:mass_sigma}) and (\ref{eq:sigma_limit}), we obtain a
final embryo mass
\begin{equation}
 M_{\rm ca} = \left[
	       \frac{2(4\alpha-5) A_{\rm ca} C_{\Sigma_{\rm s}}^{-1/4}
	       \Sigma_{\rm s,0}^{-1/4} M_{\rm iso}^{1/6}}{3 B_{\rm ca}}
\right]^{3/2(\alpha-1)}.\label{eq:Mca_ori} 
\end{equation}
Here, we assume $M_{\rm ca} \gg M_0$. 

For kilometer-sized or larger planetesimals, $Q_{\rm D}^* = Q_{\rm 0g}
\rho_{\rm p} r^{\beta_{\rm g}}$ with constants $Q_{\rm 0g}$ and
$\beta_{\rm g}$. We apply $Q_{\rm 0g} = 2.1 \,{\rm erg\, cm}^3\,{\rm
g}^{-2}$ and $\beta_{\rm g} = 1.19$ for ice \citep{benz99} and $\tilde e^2 \gg 6$, and Equation (\ref{eq:Mca_ori}) can
then be re-written as
\begin{eqnarray}
 M_{\rm ca} &=& 1.8 \times 10^{-2} 
  \left(\frac{a}{5 \,{\rm AU}}\right)^{2.8} 
  \left(\frac{m}{4\times 10^{20}\,{\rm g}}\right)^{0.63} 
  \nonumber\label{eq:Mca} 
  \\
 && \times 
  \left(\frac{Q_{\rm 0g}}{2.1 \,{\rm erg\, cm}^3\,{\rm
g}^{-2}}\right)^{1.5}
 \left(\frac{\kappa}{0.01 \,{\rm g\,cm}^{-2}}\right)^{-0.51} 
   \nonumber
  \\
 &&\times 
  \left(\frac{f_{\rm gas}\Sigma_{1}}{ 1.7 \times 10^3 \,{\rm g\,cm}^{-2}}\right)^{1.41}
M_\oplus. 
\end{eqnarray} 
Since planetesimals grow before planetesimals' fragmentation starts,
planetesimal mass $m$ is slightly larger than initial planetesimal mass
$m_0$.  Kobayashi et al. (2010) showed that planetesimals mainly
accreting onto embryos have $m = 100 m_0$.  For $m_0 \ga 10^{23}\,$g
($r_0 \ga 3 \times 10^{3}$\,km), final embryo masses exceed
$10\,M_\oplus$ at 5\,AU in a MMSN disk, but embryos cannot reach it
within a disk lifetime due to their slow growth.  The final mass $M_{\rm
ca}$ is independent of $\Sigma_{\rm s,0}$, while high $\Sigma_{\rm g,0}$
increases $M_{\rm ca}$ because gas drag highly damps $\tilde e$. For
$\Sigma_1 = 71 \, {\rm g\,cm}^{-2}$ ($10\times$MMSN), initial
planetesimals with $r_0 \ga 50$\,km can produce an embryo with
$10\,M_\oplus$ at 5\,AU.

For comparison, we also show the final mass $M_{\rm c}$ in the same
situation but neglecting the atmosphere (Kobayashi et al. 2010):
\begin{eqnarray}
 M_{\rm c} &=& 0.10 
  \left(\frac{a}{5 \,{\rm AU}}\right)^{0.63} 
  \left(\frac{m}{4\times 10^{20}\,{\rm g}}\right)^{0.48} 
\nonumber
  \\ && \times 
  \left(\frac{\ln(\Sigma_{\rm s,0}/\Sigma_{\rm s})}{4.5}\right)^{1.21} %
  \left(\frac{Q_{\rm 0g}}{2.1 \,{\rm erg\, cm}^3\,{\rm
g}^{-2}}\right)^{0.89}
      \nonumber
  \\
 &&\times 
  \left(\frac{f_{\rm gas}\Sigma_{1}}{ 1.7 \times 10^3 \,{\rm
   g\,cm}^{-2}}\right)^{1.21}
M_{\oplus}, 
\end{eqnarray}
where $\ln(\Sigma_{\rm s,0}/\Sigma_{\rm s}) \simeq 4.5$ is estimated
from Equation(\ref{eq:sigma_limit}) with $C_{\Sigma_{\rm s}}=0.1$ for
$M=0.1 M_{\oplus}$ in the MMSN model.  The collisional enhancement due
to the atmosphere is inefficient for $m = 4 \times 10^{20}$\,g;
$M_{\rm ca} < M_{\rm a}$. If $m \ga 4 \times 10^{22}\,$g, the atmosphere
contributes to embryo growth.

\subsection{Fragment Accretion
\label{sc:fragment}
}

As described above, planetesimals are ground down by collision cascade
and resulting small fragments spiral into the central star by gas drag.
In the steady state of collision cascade, the surface density of
planetesimals is much larger than that of fragments.  However, when the
grinding of planetesimals is much quicker than the removal of small
fragments by gas drag, fragments accumulate at the low-mass end of
collision cascade and determine the total mass of bodies. 
\rev{Embryos then grow through the accretion of such fragments. }

\rev{The specific impact energy between equal-sized bodies, $e^2 v_{\rm
k}^2/8$, should be much smaller than $Q_{\rm D}^*$ at the low-mass end; 
thus the typical fragments at the low-mass end have}
\begin{equation}
 e^2 v_{\rm k}^2 = C_{\rm L} Q_{\rm D}^*,\label{eq:condition_for_lowmass} 
\end{equation}
where $C_{\rm L} \sim 1$ is a constant. \rev{Although
\citet{kobayashi+10} used $C_{\rm L} = 1$ to determine the typical
fragment mass,} we apply $C_{\rm
L}=0.5$ \rev{to correct a mistake of factor 2 in their $e^2$}. 
\rev{Such small fragments feel
strong gas drag in Stokes regime; $C_{\rm D} = 5.5 c l_{\rm g} /u  r$, where $c$ is the sound velocity and $l_{\rm g} = l_{\rm
g,0} /\rho_{\rm g}$ is the mean free path of gas molecules with $l_{\rm
g,0} = 1.7\times 10^{-9} \,{\rm g\,cm}^{-2}$ \citep{adachi76}.  The
eccentricities of fragments at the low-mass end are much smaller than
$h_M$ and $\eta$, where $\eta = (v_{\rm k} - v_{\rm gas})/v_{\rm k}$ is
the deviation of the gas rotation velocity $v_{\rm gas}$ from the
Keplerian velocity. 
The dimensionless viscous stirring rate is given by 
$\langle P_{\rm VS} \rangle = \langle P_{\rm VS,low} \rangle = 73$ 
for $e \ll h_M$ \citep{Ohtsuki02} and 
the damping rate is expressed as $d e^2/dt = - 2 \eta e^2 / \tau$ for $e \ll \eta$
\citep{adachi76}. 
The equilibrium eccentricity} between stirring by embryos and damping by gas drag is
obtained as \citep{kobayashi+10}
\begin{equation}
 e^2 = \frac{h_M^3 \langle P_{\rm vs,low} \rangle \tau \Omega_{\rm
  K}}{2^{7/3} \pi {\tilde b} \eta},\label{eq:e_f} 
\end{equation}
Using Eqs. (\ref{eq:tau}), (\ref{eq:condition_for_lowmass}),
and (\ref{eq:e_f}) under the Stokes regime, we have the fragment mass
$m_{\rm f}$ at the low-mass end of collision cascade:
\begin{equation}
 m_{\rm f} = m_{\rm f0} M^{-3/2}, 
\end{equation}
where  
\begin{equation}
 m_{\rm f0} = 
  \left[
\frac{ 221 
M_* \tilde b C_{\rm
  L}Q_{\rm D}^* }{\langle P_{\rm VS,low} \rangle a^2 \Omega_{\rm K}^3}
  \frac{c}{l_{\rm g,0}} 
  \left(\frac{3}{4\pi\rho_{\rm p}}\right)^{1/3}
  \right]^{3/2}. 
\end{equation}

For $e \ll h_{\rm M}$, 
\citet{ida89} found that the dimensionless collision rate for $e
\ll h_{\rm M}$ is given
by $\langle P_{\rm col,low} \rangle = 11.3 \sqrt{\tilde R}$, where 
the coefficient is determined by \citet{inaba01}. 
Since the atmosphere effectively enhances an embryo radius for the
accretion of bodies, the collision rate 
is modified to be \citep{inaba_ikoma03}
\begin{equation}
\langle P_{\rm col} \rangle = \langle P_{\rm col,low} \rangle
 \sqrt{\frac{R_{\rm e}}{R}}.\label{eq:pcol_low} 
\end{equation}
We obtain \rev{the accretion rate of fragments by an embryo, $\dot M$,} from Eqs. (\ref{eq:dM_ori}) and (\ref{eq:pcol_low})
as 
\begin{eqnarray}
 \frac{d M}{d t} &=& A_{\rm fa} M^{43/42} \Sigma_{\rm s}^{6/7},\label{eq:dM_fra} 
\\
 A_{\rm fa} &=&\left[
  \frac{F^{1/2}  \langle P_{\rm col,low} \rangle C_{\rm acc} a^2
  \Omega_{\rm k}}{m_{\rm f0}^{1/18} (3M_*)^{2/3}} \right]^{6/7}. 
\end{eqnarray}

Fragments with $m_{\rm f}$ at the low-mass end of collision cascade 
that dominate the surface density of solids $\Sigma_{\rm s}$ 
are no longer disrupted by collisions
and drift inward by gas drag. 
\rev{The drift velocity is given by $2 \eta^2 a /\tau$ and then 
the $\Sigma_{\rm s}$-reduction rate due to the radial drift 
is expressed as $d \Sigma_{\rm s} /dt = - 2 (9/4 -q) \eta^2 \Sigma_{\rm
s}/\tau$ with the assumption of $\Sigma_{\rm s} \propto a^{-q}$. 
Since $\tau$ of fragments with $m_{\rm f}$ is determined by Equations
(\ref{eq:condition_for_lowmass}) and (\ref{eq:e_f}), we have
\citep{kobayashi+10} }
\begin{eqnarray}
 \frac{d\Sigma_{\rm s}}{dt} &=& - 
B_{\rm fa} \Sigma_{\rm s} M,\label{eq:dsigma_drift} 
\\
B_{\rm fa} &=& 
\left(\frac{9}{4}-q\right)
\frac{\langle P_{\rm vs,low} \rangle \Omega_{\rm k}
\eta v_{\rm k}^2}{2^{4/3} 3 \pi M_*C_{\rm L}\tilde b Q_{\rm D}^*}.  
\end{eqnarray}
\rev{
Since fragments are later produced by embryo growth in an outer disk, 
the radial distribution depends on time in contrast to the assumption of
$\Sigma_{\rm s} \propto a^{-q}$. Nevertheless, the effect is negligible
for embryo growth unless the atmosphere is considered
\citep{kobayashi+10}. We discuss this effect with the atmospheric
enhancement in
\S\ref{sc:simulation} and \S\ref{sc:discussion}. 
}

We can now obtain the final embryo mass $M_{\rm fa}$ 
from $\dot M$ and $\dot \Sigma_{\rm s}$ for fragment accretion, similar
to the case of planetesimal accretion. 
Integration of Equation~(\ref{eq:dM_fra}) divided by
Equation~(\ref{eq:dsigma_dt}) results in 
\begin{equation}
 M_{\rm fa} = \left(\frac{41 A_{\rm fa}}{36 B_{\rm fa}}\right)^{42/41}\Sigma_{\rm
  s,0}^{36/41}, 
\end{equation}
where we assume $M_{\rm fa} \gg M_0$ and $\Sigma_{\rm s,0} \gg \Sigma_{\rm
s}$. For $q=3/2$, we have
\begin{eqnarray}
 M_{\rm fa} &=& 0.20 
  \left(\frac{a}{5 \,{\rm AU}}\right)^{117/164} 
\left(\frac{\kappa}{0.01 \,{\rm g\,cm}^{-3}}\right)^{1/7}   
  \nonumber
  \\
&& \times  
  \left(\frac{f_{\rm ice} \Sigma_1}{30 \,{\rm
   g\,cm}^{-2}}\right)^{36/41} 
  \nonumber
  \\
&& \times  
  \left(\frac{Q_{\rm D}^*}{3.1 \times 10^6 \,{\rm erg \,
 g}^{-1}}\right)^{42/41}
  M_\oplus.\label{eq:Mfa} 
\end{eqnarray}
Here, we adopted $f_{\rm ice}$ and $\Sigma_1$ for the minimum mass solar
nebula model. 
The weak dependence of $M_{\rm fa}$ on $\kappa$ implies that 
the overestimate of $R_{\rm e}/R$ due to the power-law radial profile 
is insignificant,
as discussed in Section \ref{sc:enhance_rad}. 

For the case without an atmosphere, \citet{kobayashi+10} 
derived a final mass for the fragment accretion,  
\begin{eqnarray}
 M_{\rm f} &=& 0.14 \left(\frac{a}{5\,{\rm AU}}\right)^{3/8} 
  \left(\frac{f_{\rm ice} \Sigma_1}{30 \,{\rm g\,cm}^{-2}}\right)^{3/4} 
\nonumber
\\ && \times 
\left(\frac{Q_{\rm D}^*}{3.1 \times 10^6 \,{\rm erg \,
 g}^{-1}}\right)^{3/4} M_\oplus.\label{Mf} 
\end{eqnarray}
Eqs. (\ref{eq:Mfa}) and (\ref{Mf}) imply that the final masses increase due
to the atmosphere, but the enhancement is insignificant; $M_{\rm
fa}/M_{\rm f} \simeq 1.4$--2 for $1$--$10\times$MMSN.

\begin{figure}[htb]
\epsscale{0.6} 
\plotone{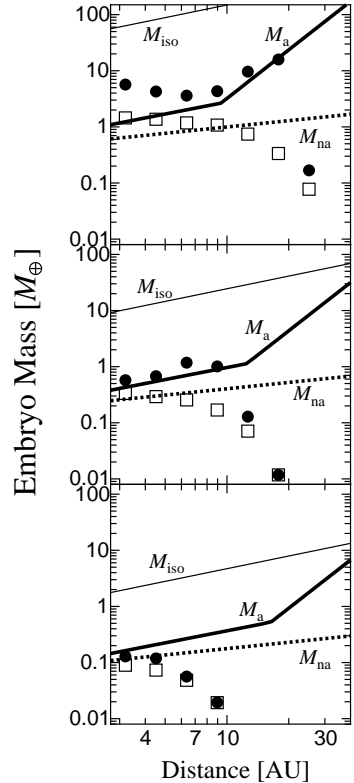} \figcaption{ 
Embryo masses with (circles) and without (squares) atmosphere after
 $10^7$ years for $m_0=4.2\times 10^{18}\,$g ($r_0=10$\,km), 
  as a function of distance form the central star. 
We set $\Sigma_1 = 71\,{\rm g \,cm}^{-2}$ (top), $\Sigma_1 = 21\,{\rm
 g \, cm}^{-2}$ (middle), and $\Sigma_1 = 7.1\,{\rm g \, cm}^{-2}$ (bottom). 
Solid lines indicate $M_{\rm a}$ which is the larger of $M_{\rm ca}$
 and $M_{\rm fa}$ for $\kappa = 0.01 {\rm cm}^2 {\rm g}^{-1}$.
Dotted lines represent 
$M_{\rm na}$ which is the larger of $M_{\rm c}$ and $M_{\rm f}$. 
Thin lines show $M_{\rm iso}$. 
\label{fig:final_mass_r10}}
\end{figure}

\begin{figure}[htb]
\epsscale{0.6} 
\plotone{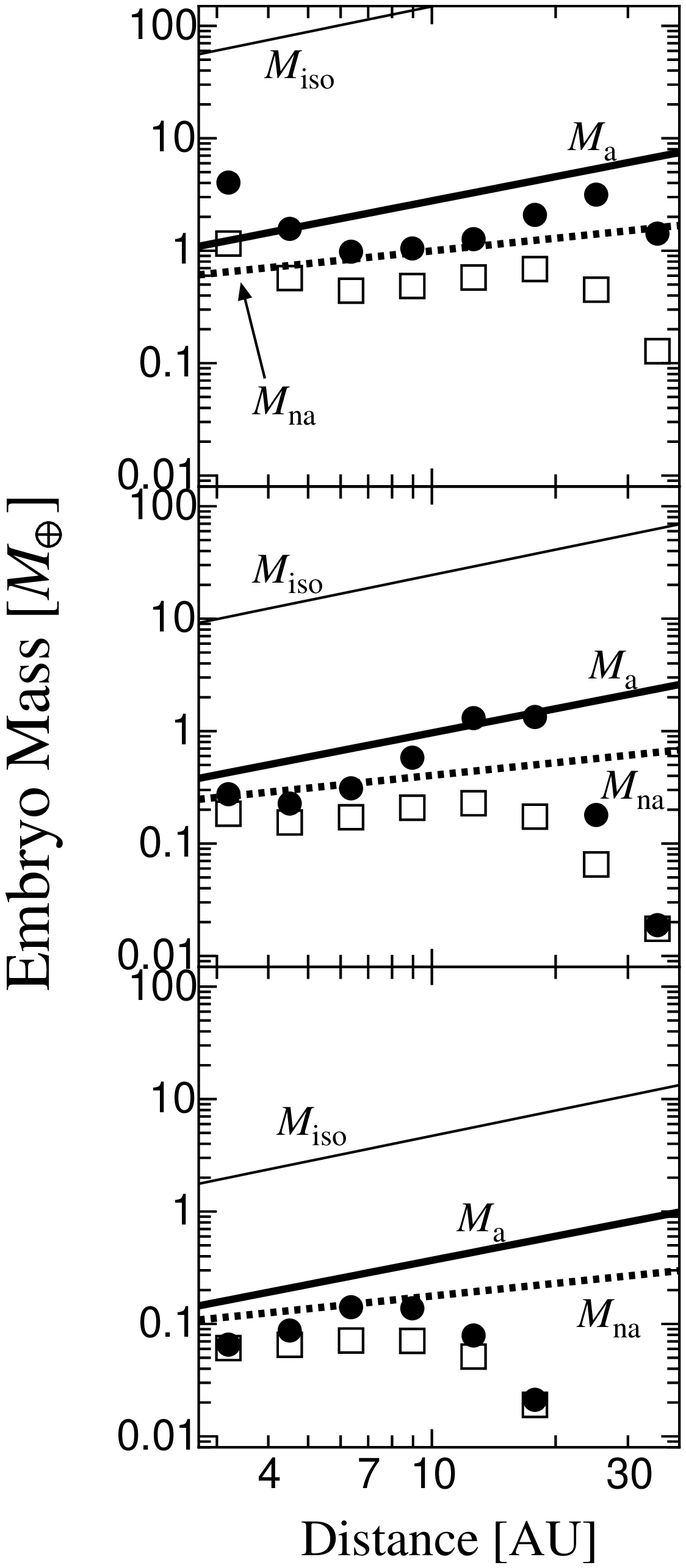} \figcaption{ 
Same as Fig.~\ref{fig:final_mass_r10}, but for $m_0 = 4.2\times
 10^{15}\,$g (radii of $1\,$km). 
\label{fig:final_mass_r1}}
\end{figure}

\begin{figure}[htb]
\epsscale{0.6} 
\plotone{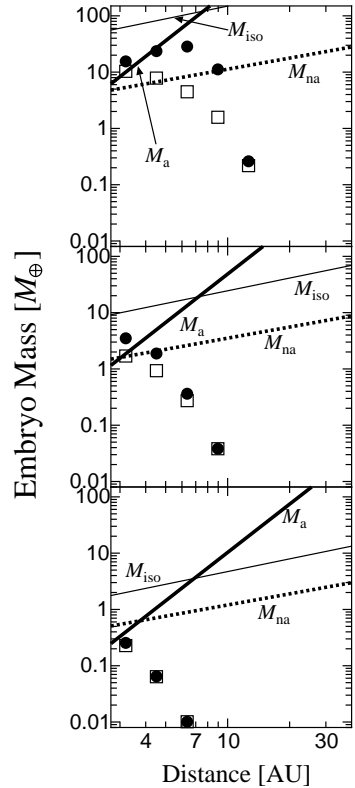} \figcaption{ 
Same as Fig.~\ref{fig:final_mass_r10}, but for $m_0 = 4.2\times
 10^{21}\,$g (radii of $100\,$km). 
\label{fig:final_mass_r100}}
\end{figure}

If we neglect the collisional enhancement due to atmosphere, the final
mass $M_{\rm na}$ is determined by the larger of $M_{\rm c}$ and $M_{\rm
f}$ (Kobayashi et al. 2010). In the case with atmosphere, a final mass
$M_{\rm a}$ is also given by the larger of $M_{\rm ca}$ and $M_{\rm
fa}$.  The final mass $M_{\rm a}$ is shown in
Figs.~\ref{fig:final_mass_r10}--\ref{fig:final_mass_r100}.  For the
initial planetesimal radius $r_0 = 10$\,km, $M_{\rm a}$ is dominated by
$M_{\rm fa}$ inside the point where the line of $M_{\rm a}$ bends in
Fig.~\ref{fig:final_mass_r10} and by $M_{\rm ca}$ outside.  The final
mass $M_{\rm a}$ is determined only by $M_{\rm fa}$ for $r_0 = 1\,$km
(Fig.~\ref{fig:final_mass_r1}) and by $M_{\rm ca}$ for $r_0=100\,$km
(Fig.~\ref{fig:final_mass_r100}) in the range of interest.

\section{NUMERICAL SIMULATION}
\label{sc:simulation}

Regarding the method of numerical simulation, we basically follow
\citet{kobayashi+10}. 
\rev{
The method of \citet{kobayashi+10} is briefly explained
here. In the calculation, a disk is divided into concentric annuli
and each annulus contains a set of mass batches.  
We set the mass ratio between the adjacent batches to 1.2, 
which can reproduce the collisional growth of bodies resulting from $N$-body
simulation without 
fragmentation \citep{kobayashi+10} and the analytical solution of mass
depletion due to collisional grinding \citep{kobayashi10}. 
The mass and velocity evolution of bodies and their radial transport
are calculated as follows. 
 \begin{itemize}
  \item[-] The mass distribution of bodies evolves through their mutual
	collisions that produce mergers and fragments. The total mass
	of fragments ejected by a single collision is given by Equation
	(\ref{eq:fragmass})  and the remnant becomes a merger. 
	   The collision rates between the bodies are calculated from the
	   formulae of \citet{inaba01}. 
  \item[-] The random velocities given by $e$ and
	   $i$ of the bodies 
	simultaneously evolve through their mutual gravitational
	interactions, gas drag, and collisional damping. The formulae of
	\citet{Ohtsuki02} are applied to describe the changing rates of $e$ and
	$i$. The gas-drag damping rates of $e$
	and $i$ are described as functions of $e$, $i$, $\eta$, and
	$\tau$ according to \citet{inaba01}. To determine $\tau$, we take into account Stokes and
	Epstein drag as well as a drag law with a quadratic dependence
	on velocity. For the collisional damping, both fragments and a
	merger resulting from a single collision have the velocity
	dispersion at the gravity center of colliding bodies. 
  \item[-] In each annulus there is a loss and gain of bodies due to
	   their inward drift. The number loss rate from an annulus is
	   given by $\int (N(m)v_{\rm drift} / \Delta a) dm$, where
	   $v_{\rm drift}$ is the drift
	velocity of bodies, $N(m) dm $ is the number of bodies with mass
	   ranging from $m$ to $m+dm$ in the annulus, 
	and $\Delta a$ is the width of the annulus. The bodies lost from
	   each annulus are added to the next inner annulus. The drift
	   velocity is given by \citep{kobayashi+10}
\begin{equation}
 v_{\rm drift} = \frac{2 a \eta}{\tau} \frac{\tilde \tau_{\rm
  stop}^2}{1+\tilde \tau_{\rm stop}^2} 
  \left[ \frac{(2E+K)^2}{9 \pi^2} e^2 +
   \frac{4}{\pi^2} i^2 + \eta^2 \right]^{1/2}, 
\end{equation}
	   where $E = 2.157$, $K = 1.211$ and the dimensionless stopping
	   time $\tilde \tau_{\rm stop} = \Omega_{\rm k} \tau /
	   (e+i+\eta)$ is adopted. 
 \end{itemize}

}

In this paper, we add a collisional enhancement
due to the atmosphere. Although the simple power-law radial density
profile of the atmosphere (Equation~(\ref{eq:atm_dens})) is used for the
derivation of final masses ($M_{\rm ca}$, $M_{\rm fa}$), the simulation
incorporates a more realistic profile provided by the formulae of
\citet{inaba_ikoma03}. The opacity of the embryo's atmosphere in their
model is given by $\kappa = \kappa_{\rm gas} + f \kappa_{\rm gr}$, where
$\kappa_{\rm gas}$ is the gas opacity, $\kappa_{\rm gr}$ is the opacity
of grains having an interstellar size distribution, and $f$ is the grain
depletion factor. Following Inaba \& Ikoma, we adopt
\begin{equation}
 \kappa = \left\{
  \begin{array}{lll} 
   \displaystyle 
    0.01 + 4 f \,{\rm cm}^2\,{\rm g}^{-1} & {\rm for} & T \le 170\,{\rm K}, 
    \\
   \displaystyle
    0.01 + 2 f \,{\rm cm}^2\,{\rm g}^{-1} & {\rm for} &  170\,{\rm K} < T \le 1700\,{\rm K}, 
    \\
   \displaystyle
    0.01\,{\rm cm}^2\,{\rm g}^{-1} & {\rm for} &  T > 1700\,{\rm K}. 
\end{array}
  \right.\label{eq:kappa}
\end{equation}
The enhancement factor $R_{\rm e}/R$ due to the atmosphere is shown in
Fig.~\ref{fig:enhanced_radius}.

We perform the \rev{simulations for embryo formation}
starting from \rev{a monodisperse mass population of} planetesimals of
mass $m_0$ and radius $r_0$ with $e = 2 i = (2 m_0/M_*)^{1/3}$ and
$\rho_{\rm p} = 1\,{\rm g\,cm}^{-3}$ around the central star of mass
$M_\sun$ with a set of eight concentric annuli at 3.2, 4.5, 6.4, 9.0,
13, 18, 25, and 35\,AU containing $\Sigma_{\rm gas}$ and $\Sigma_{\rm
s}$ for $q = 3/2$.  To compute $Q_{\rm D}^*$, we use
Equation~(\ref{eq:qd}) with $Q_{\rm 0s}=7.0 \times 10^7$\,${\rm erg}\,
{\rm g}^{-1}$, ${\beta_{\rm s}}=-0.45$, $Q_{\rm 0g}=2.1$\,erg\,${\rm
cm}^3\,{\rm g}^{-2}$, ${\beta_{\rm g}} = 1.19$, and $C_{\rm gg} = 9$
\citep{benz99,stewart09}.  We artificially apply the gas surface density
evolution in the form $\Sigma_{\rm gas} = \Sigma_{\rm
gas,0}\exp(-t/T_{\rm gas,dep})$, where $T_{\rm gas,dep}$ is the gas
depletion timescale, which we set to $T_{\rm gas,dep} = 10^7$ years.
Assuming a constant $\Sigma_{\rm gas}$ gives almost the same results for
final embryo masses, because we consider time spans $t \leq T_{\rm
gas,dep}$.

\begin{figure}[htb]
\epsscale{0.6} \plotone{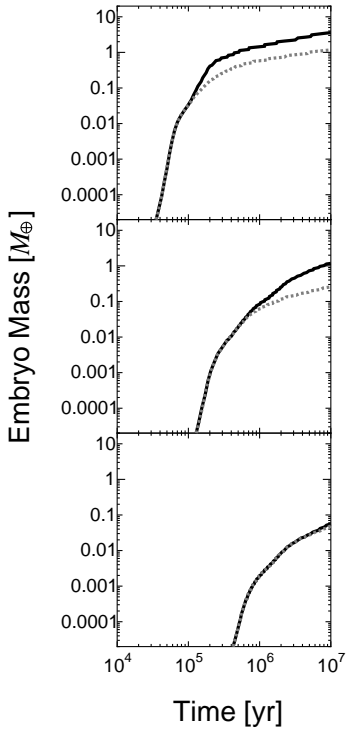} \figcaption{ Evolution of
embryo mass at 6.4\,AU with $m_0 = 4.2 \times 10^{18}$\,g ($r_0 =
10$\,km) for $\Sigma_0 = 71\,{\rm g \,cm}^{-2}$ ($10\times$MMSN; top),
$21\,{\rm g \, cm}^{-2}$ ($3\times$MMSN; middle), and $7.1\,{\rm g \,
cm}^{-2}$ (MMSN; bottom). Solid lines show the case with atmosphere and
dotted lines represent the result without atmosphere.
\label{fig:comp_growth}}
\end{figure}

Fig.~\ref{fig:comp_growth} shows the \rev{embryo-mass evolution} at
6.4\,AU for $f=0.01$.  
 \rev{Runaway growth initially occurs;
embryo mass exponentially grows with time during the stage. The runaway-growth timescale is proportional to $r_0/\Sigma_{\rm s,0}$
\citep{ormel10a,ormel10b}. When the embryo masses exceed
0.001-0.01$M_\oplus$, oligarchic growth starts. 
Since massive embryos dynamically excite planetesimals, 
the reduction of planetesimals due to 
collisional fragmentation stalls the embryo growth \citep{kobayashi+10}. 
For $\Sigma_0 = 7.1\,{\rm g\,cm}^{-2}$ (MMSN), 
the fragmentation limits the final mass to about Mars mass ($\sim 0.1 M_\oplus$) and the atmosphere is insignificant. 
}
Once embryo masses exceed the Mars
mass, atmosphere substantially accelerates the embryo growth. For
$\Sigma_0 \geq 21\,{\rm g\,cm}^{-2}$ ($3\times$MMSN), the atmosphere
leads to further embryo growth. Nevertheless, embryos finally attain
asymptotic masses.

\begin{figure}[htb]
\epsscale{0.9} \plotone{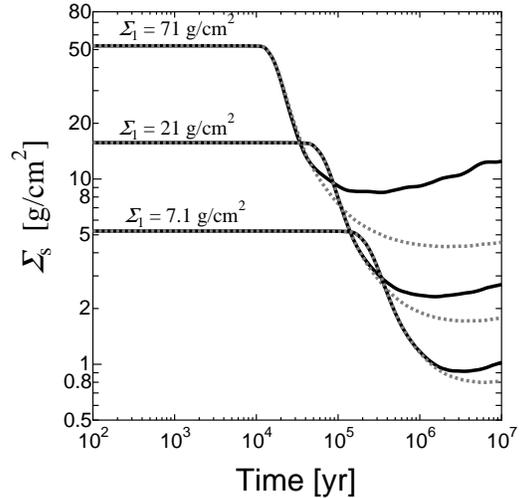} \figcaption{ 
The solid surface density evolution at 3.2\,AU. 
\label{fig:comp_r10_sig}}
\end{figure}

Results for these simulations are summarised in
Fig.~\ref{fig:final_mass_r10}, where the embryo masses after $10^7$
years are compared to analytical formulae for final embryo masses.
Embryo masses finally reach $M_{\rm a}$ inside 5\,AU ($\Sigma_0 =
7.1\,{\rm g\,cm}^{-2}$), 10\,AU ($\Sigma_0 = 21\,{\rm g\,cm}^{-2}$), and
20\,AU ($\Sigma_0 = 71\,{\rm g\,cm}^{-2}$).  However, embryos exceed
$M_{\rm a}$ inside 5\,AU for $\Sigma_0 = 71\,{\rm g\,cm}^{-2}$.  This
excess comes from the embryo growth through collisional accretion with
bodies drifting from outside, which effect we did not consider in the
analysis described in Section \ref{sc:final_mass}.  To confirm the
contribution from drifting bodies, we show the surface density evolution
in Fig.~\ref{fig:comp_r10_sig}. For $\Sigma_0 = 71\,{\rm g\,cm}^{-2}$,
the surface density of solids increases after
$2\times10^5$\,years. Since the drift timescale shortens inward, bodies
from outside cannot raise the surface density unless embryos accrete
them. Therefore, the increase in the surface density implies that embryo
grows through the accretion of such bodies.

The initial mass $m_0$ of planetesimals in the simulations depends on
their formation process, which is not well understood yet.  We perform
the embryo growth starting from different $m_0$
(Figs.~\ref{fig:final_mass_r1} and \ref{fig:final_mass_r100}).  Small
planetesimals are relatively easily fragmented due to low $Q_{\rm D}^*$
and quickly ground down to the low-mass end of collision cascade.  The
resulting fragments with low $e$ actively accrete onto embryos.  For
$m_0 = 4.2 \times 10^{15}\,$g ($r_0=1$\,km), embryos can reach a final
mass $M_{\rm a}$ in a relatively wide region inside 10\,AU (MMSN),
20\,AU ($3\times$MMSN), and 30\,AU ($10\times$MMSN). On the other hand,
large initial planetesimals delay the runaway growth of embryos
\citep{ormel10a,ormel10b} and the following oligarchic growth is also
slower than that for small planetesimals because embryos mainly accrete
original planetesimals rather than fragments with low $e$. \rev{For $r_0
= 100\,$km, }embryos
attain the final masses only inside 4\,AU for $3\times$MMSN and inside
6\,AU for $10\times$MMSN, and embryos cannot reach final masses beyond
2.7\,AU in the MMSN disk.  In addition, small bodies drifting from
outside are effectively captured by embryos and thereby embryos exceed
final masses $M_{\rm a}$ inside 4\,AU for $10\times$MMSN.

In the case without an atmosphere, initially larger planetesimals can
form massive embryos. Since large planetesimals delay embryo growth,
embryos made from 100\,km-sized initial planetesimals can reach
$10\,M_\oplus$ but the location is only inside 3--4\,AU even for
$10\times$MMSN \citep{kobayashi+10}. The case with the atmosphere
shows a similar dependence of the final embryo masses on initial
planetesimal mass. However, since the atmosphere accelerates embryo
growth, embryos larger than $10\,M_\oplus$ are produced inside 8--9\,AU
of a $10\times$MMSN disk with 100\,km-sized initial planetesimals.

\begin{figure}[htb]
\epsscale{0.9} \plotone{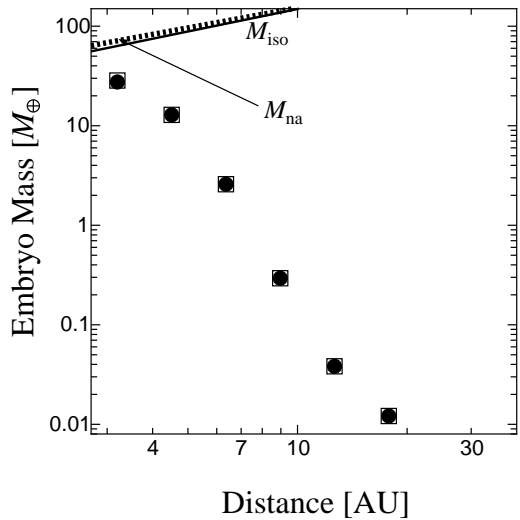} \figcaption{ 
Same as Fig.~\ref{fig:final_mass_r10}, but for $\Sigma_1 = 71\,{\rm
 g \, cm}^{-2}$ with $m_0 = 4.2\times
 10^{24}\,$g ($r_0 = 1000\,$km). 
\rev{The final mass $M_{\rm a}$ with
 atmosphere is estimated to be larger than $200 M_\oplus$. }
\label{fig:final_mass_r1000}}
\end{figure}

While the final masses of embryos exceed $10M_\oplus$ for large initial
planetesimals of $r_0\ga 100$\,km, embryos must reach the critical core
mass within the disk lifetime $T_{\rm gas,dep}$ to form gas giant
planets.  The growth timescale is estimated to be $M/\dot M$, where
$\dot M$ is given by Equation~(\ref{eq:dM_pla}).  The critical distance
$a_{\rm c}$ inside which embryos can reach $10M_{\oplus}$ is
approximately obtained from the condition $M /\dot M < T_{\rm gas,dep}$
with $M=10M_{\oplus}$,
\begin{eqnarray}
 a_{\rm c} &=& 9.6 \left(\frac{T_{\rm dep}}{10^7 {\rm
		  years}}\right)^{20/39}
 \left(\frac{\Sigma_1}{71\,{\rm g\,cm}^{-2}}\right)^{23/39}
 \nonumber \\ && \times
 \left(\frac{r_0}{100\,{\rm km}}\right)^{-1/3}\,{\rm AU},\label{eq:crit_dis}
\end{eqnarray}
where we adopt $m=100\,m_0$ and $q=3/2$.  For $r_0 =100$\,km, the
massive disk with $\Sigma_1 \ga 70\,{\rm g\,cm}^{-2}$ can form such
large embryos around $10$\,AU.  In addition, we estimate $a_{\rm c} \sim
5\,$AU from Equation~(\ref{eq:crit_dis}) for a $10\times$MMSN disk with
$r_0 = 10^3$\,km. Indeed, the simulation with $\Sigma_1 = 71\,{\rm
g \,cm}^{-2}$ and $r_0 = 100\,$km shows embryos cannot reach
$10M_\oplus$ beyond 5\,AU (see Fig.~\ref{fig:final_mass_r1000}).
Therefore, the condition of $10\times$MMSN with $r_0 \sim 100$\,km is
necessary to form gas giants around 10\,AU.

\begin{figure}[htb]
\epsscale{0.6} 
\plotone{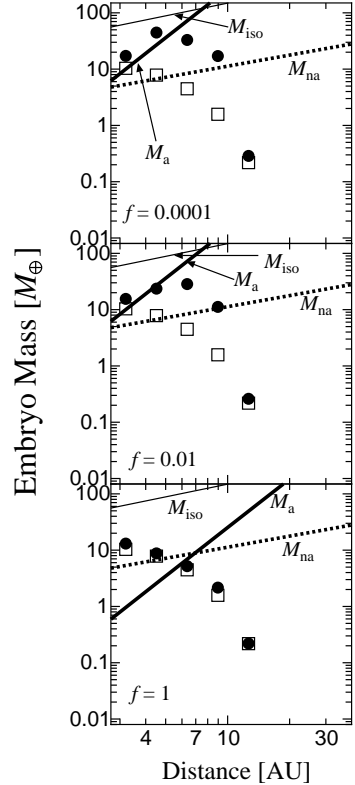} \figcaption{ 
The final embryo masses for $f=0.0001$ (top), 0.01 (middle) and 1 (bottom),
 starting from $10\times$MMSN with $m_0 =
 4.2\times 10^{21}\,$g ($r_0 = 100\,$km). 
Lines and symbols
are the same as in Fig.~\ref{fig:final_mass_r10},
but we apply $\kappa = 1 {\rm cm}^2 \, {\rm g}^{-1}$ to
derive $M_{\rm a}$ for $f=1$. 
\label{fig:comp_f}}
\end{figure}

We also give a constraint on $f$.  For $f \la 0.01$, a final mass is
almost independent of $f$ (see Fig.~\ref{fig:comp_f}). This is because
the gas opacity dominates over the grain opacity (see
Equation~(\ref{eq:kappa})).  For $f=1$, embryos at 3--4\,AU become
larger due to the capture of bodies drifting from outside, while final
embryo masses in the outer disk are similar to the case without
atmosphere.  The condition of $f \la 0.01$ is therefore necessary for
gas giant formation in the region 5--$10$\,AU and such low $f$ is
acceptable; the depletion factor $f$ should be much smaller than unity
after planetesimal formation.  In addition, a low-opacity atmosphere
reduces the critical core mass \citep{mizuno80,ikoma00,hori10}. 

\section{DISCUSSION}
\label{sc:discussion}

We derived final embryo masses analytically and numerically.  They agree
with each other quite well in the inner disk where the embryo formation
timescale is shorter than the nebula lifetime ($\sim 10^7\,$years).  The
analytical formula for final masses $M_{\rm a}$ implies that initial
planetesimal radii should be larger than about $3 \times 10^{3}$\,km to
form embryos with $10\,M_\oplus$ at 5\,AU in a MMSN disk. However, the
critical distance $a_{\rm c}$ inside which embryos reach $10\,M_\oplus$
within $10^7$ years (Equation~(\ref{eq:crit_dis})) is estimated to be
much smaller than 5\,AU; a massive disk is likely to form gas giant
planets.  Embryos inside 5\,AU of a $\sim 10\times$MMSN disk exceed
final embryo masses $M_{\rm a}$ due to the accretion of small bodies
drifting from outside.  In spite of such further growth, embryos
starting from small planetesimals cannot reach the critical core mass
$\sim 10\,M_\oplus$.  In addition, further growth is insignificant
beyond 5\,AU.  The formulae for $M_{\rm a}$ and $a_{\rm c}$ suggest that
initial planetesimals with $r_0 \simeq 50$--700\,km are necessary for
embryos to reach $10\,M_\oplus$ at 5\,AU in the $10\times$MMSN disk.

Inaba et al. (2003) performed similar simulations incorporating
collisional fragmentation and enhancement due to the embryo's atmosphere
and showed a planetary core with $M>10M_\oplus$ could be produced around
5\,AU with $m_0 = 4.2 \times 10^{18}$\,g ($r_0=10$\,km) for
$10\times$MMSN.  In our simulation, embryos cannot reach $10M_\oplus$
under this condition and larger planetesimals are necessary to form such
massive embryos beyond 5\,AU.  As \citet{kobayashi10} discussed,
\citet{williams} underestimated the total ejecta mass produced by a
single collision for cratering; Inaba et al.\ adopted the fragmentation
model similar to theirs that \citet{wetherill93} developed (see
Fig.~\ref{fig:me}).  Erosive collisions shorten the depletion time of
10km-sized {planetesimals in} collision cascade by a factor of 4--5
\citep{kobayashi10} and hence reduce final embryo masses.  As seen from
Eqs.~(\ref{eq:Mca}) and (\ref{eq:Mfa}), final embryo masses $M_{\rm
ca}$, $M_{\rm fa}$ increase with $Q_{\rm D}^*$; the results of Inaba et
al. correspond to embryo masses for higher $Q_{\rm D}^*$.  Although we
and Inaba et al.  applied $Q_{\rm D}^*$ provided by \citet{benz99},
porous bodies with $r \la 10$\,km may have much lower $Q_{\rm D}^*$
\citep[e.g.,][]{stewart09,machii}.  For initial planetesimals with radii
$\ga 100\,$km, $Q_{\rm D}^*$ of slightly larger bodies determines final
embryo masses and is almost entirely determined by the gravitational
binding energy; the uncertainty from their structure would be minor.
Therefore, such large planetesimals are possible to produce cores
for gas giant planets.


The mechanisms of planetesimal formation are highly debated but, despite
intensive effort, remain fairly unknown. The formation through
collisional coagulation in which dust smoothly grows to planetesimals
with $r_0 \sim 1$\, km face barriers: meter-sized objects should be lost
to the central star as a result of gas drag 
\citep{weidenschilling77,brauer08}, and further agglomeration of cm-sized objects upon
collision is problematic because of collisional bouncing 
\citep{guttler10,zsom10}. 
Moreover, the electric repulsion may stop
growth of smaller objects \citep{okuzumi09}.  A new scenario that
allows one to overcome the barriers has been proposed recently:
self-gravity of small particles accumulating in turbulent structures of
gaseous disks forms large planetesimals of the order of $100$\,km 
\citep{johansen07,cuzzi08}. Not only do such large
planetesimals produce planetary cores exceeding the critical core mass
to form gas giant planets, they may also be consistent with
properties of minor bodies in the solar system. Indeed, the initial
planetesimals should be larger than 100\,km to reproduce the mass
distribution of asteroids in the main belt \citep{morbidelli09}. 


For large planetesimals, a final embryo mass given by $M_{\rm ca}$ is
large enough to start core accretion, while embryo growth is slow.  If
the radial slope of surface density $q=3/2$ like the MMSN model, a
massive disk with $10\times$MMSN is necessary for embryos to reach the
final mass around 10\,AU.  However, observations of protoplanetary disks
infer their relatively flatter radial distributions over several hundred
AU \citep[e.g.,][]{kitamura02}. 
In such a disk, dust grains accumulate
in an inner disk due to radial drift during their growth, which
increases the solid surface density in the inner disk \citep{brauer08}. 
The enhancement of solid surface density accelerates embryo
growth and hence embryos may achieve the critical core mass in less
massive disks.


To form gas giants via core accretion, rapid gas accretion onto a core
with $\sim 10$ Earth masses must occur prior to gas depletion. However,
these cores migrate inward due to their exchange of angular momentum
with the surrounding gas (Type I). From linear analysis, the
characteristic orbital decay time of Earth-mass cores at several AU in
the MMSN model is about 1\,Myr \citep{tanaka02}. Several processes
to delay the timescale of Type I migration have been pointed out, for
example, disk surface density transitions \citep{masset06b}, 
intrinsic turbulence \citep{nelson04}, and hydrodynamic
feedback \citep{masset06a}. There is still uncertainty about this
estimate of the migration time.  Indeed, the distribution consistent
with observations of exoplanets can be reproduced only if the timescale
of the type I migration is at least an order of magnitude longer than
that derived from the linear analysis \citep{ida_lin08}.  We should
also investigate the strength of such migration for the survival of
cores of gas giant planets in our future work.



\section{SUMMARY}
\label{sc:summary}

In this paper, we investigate the growth of planetary embryos by taking
into account, among others, two effects that are of major importance.
One of them is collisional fragmentation of 
planetesimals, which is induced by their gravitational interaction with
planetary cores. Another effect is an enhancement of collisional cross
section of a growing embryo by a tenuous atmosphere of nebular gas, 
which becomes substantial when 
an embryo has reached about a Mars mass.

The main results are summarized as follows.
	   
\begin{itemize}
 \item[1.] If the atmosphere is not taken into account, collisional 
           fragmentation suppresses planetary embryo growth
	   substantially.
           As a result, embryos cannot reach the critical core mass of
           $\sim 10 M_\oplus$ needed to trigger rapid gas accretion to form
           gas giants. The final masses
	   are about Mars mass in a MMSN disk \citep{kobayashi+10}.
           Embryo's atmosphere accelerates the
	   embryo growth and may increase the final embryo mass by
           up to a factor of ten.

 \item[2.]  Planetary embryos attain their final masses asymptotically.
           We have derived the final mass analytically.
           The final mass of an embryo is predicted to be
           the larger of $M_{\rm ca}$ and $M_{\rm fa}$,
           which are given by Eqs.~(\ref{eq:Mca}) and (\ref{eq:Mfa}),
           respectively.
           These final masses are in good agreement with the
	   results of statistical simulations. 

  \item[3.] 
	   Our solution indicates that an initial planetesimal radius $r_0 \ga
	   3\times 10^3$\,km is necessary to form a planetary core with
	   $10\,M_\oplus$ at 5\,AU in a MMSN disk. However, such initially
           large planetesimals delay embryo growth; a massive disk is
	   required to produce massive cores within a disk lifetime. The
	   analytical solution for the final mass and the embryo
	   formation time show that planetesimals with an initial
           radius of $r_0 \simeq 50$--700\,km are likely to produce such
	   a large planetary core within a disk lifetime at 5\,AU for
	   $10\times$MMSN.

  \item[4.]
           The embryo growth depends on the disk mass, initial
	   planetesimal sizes, and the opacity of atmosphere. We have
	   performed statistical simulations to calculate the final embryo 
           masses over a broad range of parameters. We took the surface 
           density of solids at $1$~AU in the range
	   of $\Sigma_1 = 7.1$--$71{\rm g\,cm}^{-2}$ (1--$10\times$MMSN),
	   initial planetesimal radius $r_0 = 1$--1000\,km, and 
           the grain depletion factor $f$ in planetary
	   atmosphere between $f=10^{-4}$--1.
           We found that planetary embryos can exceed
	   $10M_\oplus$ within 8-9\,AU for $10\times$MMSN, $r_0 =
	   100$\,km, and $f\leq 0.01$.  Other sets of parameters cannot
	   produce massive cores at 5--10\,AU.  For example, embryo's mass
	   can reach $6\,M_\oplus$ for $r_0 = 10$\,km only inside
	   4\,AU. Therefore, we conclude that a massive disk ($\sim
	   10\times$MMSN) with $r_0 \sim 100$\,km and $f\la 0.01$ is
	   necessary to form gas giant planets around 5--10\,AU. 
           This condition for large embryo formation
	   is independent of the material strength and/or structure of
	   bodies, because $Q_{\rm D}^*$ of 100km-sized or larger bodies 
           is largely determined by their self-gravity.
\end{itemize}

\vspace{2cm}
\rev{
We thank Chris Ormel for helpful discussions and the reviewer, John Chambers, for
useful comments on the manuscript. 
}

\end{document}